\documentclass[preprint]{aastex}
\slugcomment{Accepted by the PASP: 25 October 2000}

\shorttitle{Up-the-Ramp Sampling on IR Detectors}
\shortauthors{Offenberg, et al.}

\newcommand{\etal}{et~al.}
\newcommand{\mm}{\mu{\rm m}}
\begin{document}

\hfuzz=10pt \overfullrule=0pt
\pretolerance=10000
\pretolerance=8000	

\title{Validation of Up-the-Ramp Sampling with Cosmic Ray Rejection on IR Detectors}

\author{
J.\ D.\ Offenberg,\altaffilmark{1,2,3} 
D.\ J.\ Fixsen,\altaffilmark{1,2} 
B.\ J.\ Rauscher,\altaffilmark{3}
W.\ J.\ Forrest,\altaffilmark{4}
R.\ J.\ Hanisch,\altaffilmark{3} 
J.\ C.\ Mather,\altaffilmark{2}
M.\ E.\ McKelvey,\altaffilmark{5}
R.\ E.\ McMurray\ Jr,\altaffilmark{5}
M.\ A.\ Nieto-Santisteban,\altaffilmark{3}
J.\ L.\ Pipher,\altaffilmark{4}
R.\ Sengupta,\altaffilmark{1} and
H.\ S.\ Stockman\altaffilmark{3}
}
\altaffiltext{1}{Raytheon ITSS, 4500 Forbes Blvd, Lanham MD 20706, Joel.D.Offenberg.1@gsfc.nasa.gov}
\altaffiltext{2}{Laboratory for Astronomy and Solar Physics, NASA's Goddard Space Flight Center, Greenbelt MD 20771}
\altaffiltext{3}{Space Telescope Science Institute, 3700 San Martin Dr, Baltimore, MD 21218}
\altaffiltext{4}{University of Rochester, Dept. Of Physics and Astronomy, Rochester NY 14627}
\altaffiltext{5}{NASA's Ames Research Center, Moffett Field CA 94035}


\begin{abstract}

We examine cosmic ray rejection methodology on data collected from
InSb and Si:As detectors.  The application of an Up-the-Ramp sampling
technique with cosmic ray identification and mitigation is the focus
of this study.  This technique is valuable for space-based
observatories which are exposed to high-radiation environments.  We
validate the Up-the-Ramp approach on radiation-test data sets with
InSb and Si:As detectors which were generated for SIRTF.  The
Up-the-Ramp sampling method studied in this paper is over 99.9\%
effective at removing cosmic rays and preserves the structure and
photometric quality of the image to well within the measurement error.

\end{abstract}


\section{Introduction}\label{sec:intro}

The effects of radiation and cosmic rays can be a formidable source of
data loss for a space-based observatory.  The authors have been
studying the question of cosmic ray identification and mitigation in
the context of processing data for the Next Generation Space Telescope
(NGST).  The deep space, high-radiation environment proposed for the
NGST (Stockman, 1997) and the long observing times needed to complete
some of the NGST Design Reference Mission programs suggest that
careful planning for cosmic ray mitigation is needed (Stockman \etal,
1998).  Although this study was motivated by the NGST requirements,
the methods and results presented here are not specific to the NGST
and may be applied to many instruments and observatories.

When a cosmic ray\footnote{We use the term ``cosmic ray'' to refer to
a charged particle which injects spurious signal on the
photo-conducting layer of the detector.} impacts a detector, an
undetermined amount of charge is deposited in the photo-conducting
layer.  When that happens, any unrecorded information stored in the
detector at that location is lost and cannot be recovered.  In a
high-radiation environment (such as deep space), the data loss due to
cosmic ray events in the detector may impose limits on observation
parameters, such as maximum integration time, which in turn will
limit---or prevent---some science programs.

Several solutions to the problem of identifying and removing cosmic
rays exist.  The most straightforward approach is to integrate
multiple times, filter the outliers (pixels in an array with signal
different from an expectation value generated from the full image
set), and co-add the resulting images.  We have additional options if
the detector can be sampled non-destructively (i.e. the detector can
be read without being reset), a feature that is available on some
current and future detectors (e.g. Fowler \& Gatley, 1990; Garnett \&
Forrest, 1993; Fanson, \etal, 1998), including those being studied
here.

We evaluate and validate Up-the-Ramp sampling with on-the-fly cosmic
ray identification and mitigation, which is described in detail by
Fixsen, \etal\ (2000).  In this method, the detector is sampled
non-destructively at uniform intervals, resulting in a set of reads
following the accumulating charge in the detector over time.  The
signal is measured as the slope of the accumulating charge ``ramp,''
and cosmic rays and similar glitches can be identified and discarded
from the signal measurement.  Several concerns about this method
exist, namely that the method is too compute-intensive, that it
requires detailed calibration with the detector and the approach is
too simplistic for the complicated processes involved.  Computation
requirements are discussed by Fixsen, \etal\ \ This paper will
validate the Up-the-Ramp method without detailed calibration or
complicated models.  

The most commonly-discussed alternative to Up-the-Ramp sampling is
Fowler sampling (Fowler \& Gatley, 1990), in which multiple samples
are taken at the start and at the end of an observation to effectively
measure the difference $N$ times, where $N$ is the number of
sample-pairs.  Fowler sampling, however, does not provide an
opportunity to identify and remove cosmic rays from an observation.
In addition, in the read-noise limit, Up-the-Ramp sampling provides
modestly ($\sim6\%$) higher signal-to-noise than Fowler sampling
(Garnett \& Forrest, 1993).  For a detailed comparison of
signal-to-noise for Fowler and Up-the-Ramp sampling, see
Appendix~\ref{App:SNR}.

We utilize data sets from radiation tests which were performed during
the design and construction of the SIRTF space-based observatory.  The
data sets are from InSb and Si:As infrared detectors; we discuss each
set individually.  We apply the Up-the-Ramp sampling algorithm
presented by Fixsen, \etal\ and examine the results.  We 
validate the Up-the-Ramp approach as a low-cost solution to mitigating
cosmic rays and demonstrate the quality of the resulting data.

Section~\ref{sec:algorithm} is a summary description of the
Up-the-Ramp processing algorithm.  Section~\ref{sec:interaction}
provides a description of the interactions between charged particles
and detectors, with particular emphasis on the effects of cross-talk
in the detectors.  Section~\ref{sec:InSbData} describes the InSb data
set; Section~\ref{sec:SiAsData} describes the Si:As data set.  We
discuss the results of the algorithm as applied to the InSb data set
in Section~\ref{sec:InSbDiscussion}.  Section~\ref{sec:SiAsDiscussion}
discusses the results from the Si:As data.  The main text concludes in
Section~\ref{sec:conclusion}.  A detailed discourse on signal-to-noise
when considering cosmic rays is in the Appendix.

\section{Up-the-Ramp Processing Algorithm}\label{sec:algorithm}

The details of the Up-the-Ramp algorithm can be found in Fixsen,
\etal\ (2000).  A shorter version is presented in Offenberg, \etal\
(1999), although the algorithm has been revised since that writing.
We provide a summary description of the method here.

The Up-the-Ramp algorithm assumes a non-destructive set of $N$ samples
for an integration.  The general approach does not require the samples
be spaced uniformly in time, but it will be most efficient if they are
and the implementation used in this study does assume uniform sampling.
When processing these data, we process each pixel individually;
although there are indications that a radiation event on the detector
will affect neighboring pixels via electronic cross-talk (see
Section~\ref{sec:interaction}), there is no attempt made to ``impugn
neighbors'' of pixels suspected of being impacted by a cosmic ray.

We first identify and remove saturated pixels.  To estimate the
signal per unit time, we need to discard the samples when the
charge-well is full and no data can be collected; in practice there is
a region of non-linear response before the detector charge-well is
full which should also be discarded.  We identify saturated pixels by
defining a cut-off value for the upper limit of the linear region of
the detector's response curve.  The samples in an observation sequence
are compared to this cut-off from the last observation to the first,
stopping when a non-saturated sample is found.  The saturated samples
are discarded from further computation.

Next, we search for cosmic rays and other glitches.  We start with a
signal estimate; a simple estimate is the mean signal accumulated
during one sample time, $s=(D_N - D_0)/N$.  We first seek the worst
point, which is measured as the sample $i$ with the maximum value $X_i
= |D_i~-~D_{i-1}-s|$.  We then examine the series by comparing the maximum
value of $X_{max}$ found to $\sigma$, which is the expected noise based
on the known read-noise and the photon shot-noise determined by the
signal estimate $s$.  If $X_{max}~>~a\sigma$, (where $a$ is a tunable
cosmic-ray threshold), $D_i$ represents a glitch.  $D_i$ is then
discarded from the data set, $s$ is updated, and the search for cosmic
rays is repeated.  We stop when the worst outlier is within the bounds
of acceptable variation.

There are several points that need to be made about the algorithm
described in the previous paragraph.  We seek cosmic ray glitches in
both the positive and negative sense for several reasons.  Although
cosmic rays will normally inject (not subtract) charge to the
detector, negative-sense glitches can occur as the result of impacts
on the electronics.  Also, the statistical test used to identify
cosmic rays will be subject to false-positive identificiations---to
avoid biasing the data, we must discard such outliers in both
directions.  We reject only one outlier on each pass as we have found
that procedure to be the most robust in the case of multiple cosmic
rays.

We then fit the remaining data to a line using a weighted
least-squares fit---the figure of interest is the signal per unit
time, which is the slope of this line.  The weights are determined by
the signal-to-noise estimate, based on the signal estimate $s$
computed earlier and the read noise of the detector; in low
signal-to-noise cases, the data points are weighted evenly for the
fit.  In high signal-to-noise cases, the end-points are weighted more
heavily than the middle-points.  To speed up the algorithm, we
precompute the coefficients to produce this weighted fit for a set of
signal-to-noise values, and select the fit corresponding to the
highest signal-to-noise which is less than the measured value.  This
approximation is less than optimal, but is very close to optimal with
as few as 8 S/N values and does not bias the results.  Underestimating
the noise in the data is a much worse case than overestimating the
noise, so we consistently choose the weighting to overestimate the
noise, corresponding to lower signal-to-noise ratio (Fixsen, \etal,
2000).

In the case where a sequence is broken up by one or more cosmic rays,
the slope is computed for each segment uninterrupted by a cosmic ray,
saturated or other glitch sample.  The slopes are then combined using
a weighted average, where each slope is optimally weighted.

We set the noise and detector saturation levels to appropriate values
for the data sets being processed, but did not otherwise alter the
algorithm or tune the cosmic ray detection algorithm parameters.  In
particular, the threshold for identifying a cosmic ray in the tests
described here was $4.5\sigma$, the optimum threshold found by
Fixsen, \etal\ for their test case.

\section{Cosmic Ray/Detector Interaction}\label{sec:interaction}

When an energetic cosmic ray interacts with an IR detector, its main
effect is to excite charge in the photo-conducting layer of the detector,
contributing signal to the detector in the region impacted.  The
precise amount of charge injected is effectively random with some
probability distribution, as it depends on several factors (e.g. the
energy of an individual cosmic ray) which are either random or which
can not be recovered {\it a posteriori}.  In short, an energetic
cosmic ray injects an unknowable amount of energy into the
detector---this, in effect, destroys the information recorded in the
impacted pixel(s) since the last measurement.

A cosmic ray interacting with the detector can have effects which
persist for a significant time after the initial impact.  For example,
if a cosmic ray liberates a very large amount of charge from the
detector's photo-conducting layer in a short time, it may take a
measurable time while electrons repopulate the detector material,
during which time the gain in that pixel might be flat, or varying
over time.  Note that this and similar effects might persist for
seconds or minutes---given sufficient recovery time, the detector will
function as before the cosmic ray hit in most cases.\footnote{Over
time, repeated particle impacts and radiation exposure will
permanently damage the detector, but the long-term degradation from a
single particle impact should not be measurable.}  The net effect of
these persistent effects is that data collected at a detector pixel
element might be invalid for a time after a cosmic ray hit.  This can
be accommodated by modifying the data--fitting algorithm to ignore a
set number of samples after a cosmic ray detection.  For this study,
we assume that persistent effects in the InSb and the Si:As arrays are
small and can be ignored.

\begin{figure}
\plotone{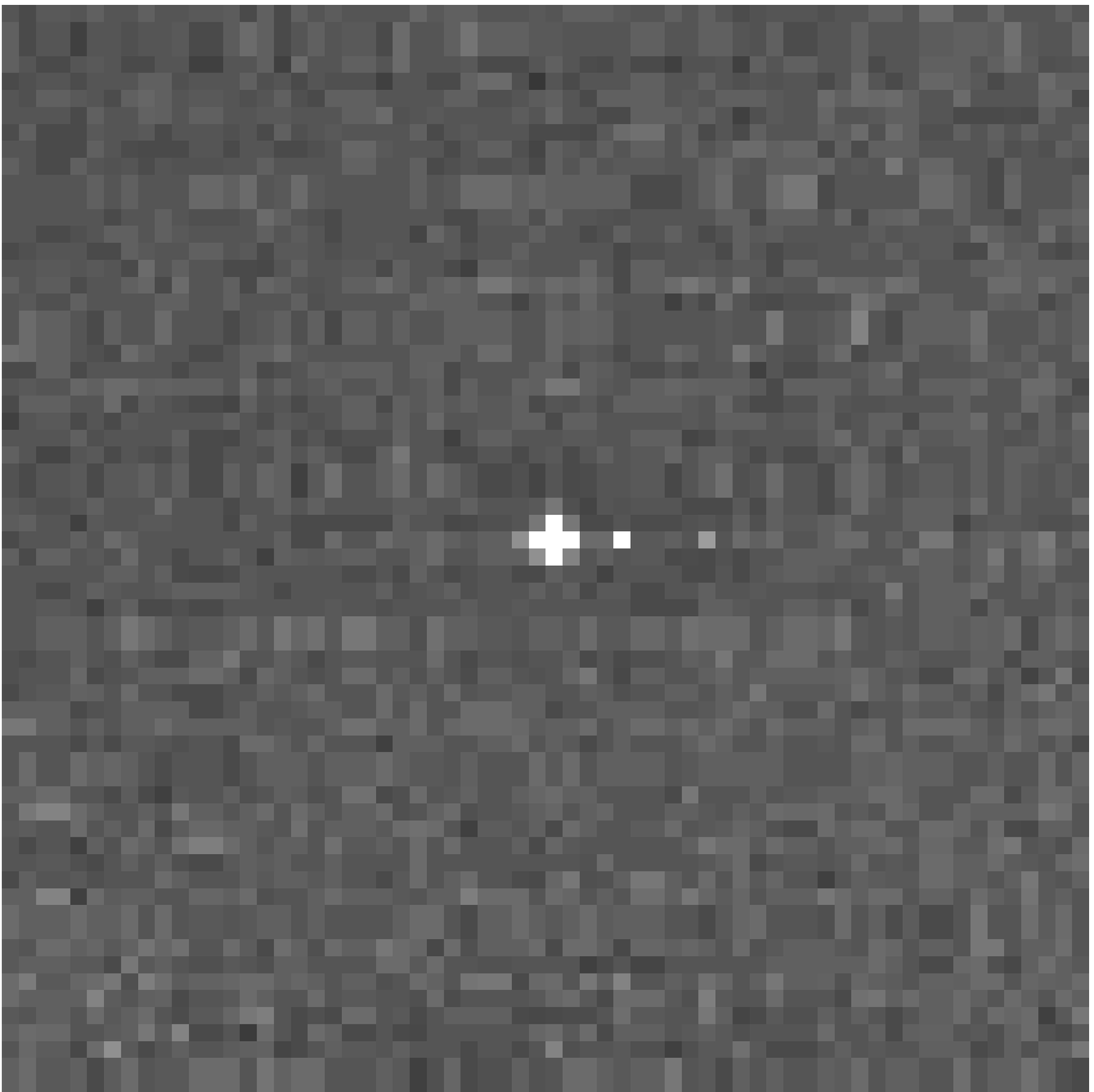}
\caption{Stacking a large number, $\sim$100, of pixel
proton hits from the InSb data set reveals cross-talk both in the
multiplexer and in the detector array. In time, the hot pixel 4 pixels
to the right of the primary hit was the next pixel read out---this hot
pixel is referred to as the ``MUX bleed.''  Moreover, the hit is not
confined to one pixel. Rather, it spreads to its neighbors. One well
understood mechanism governing this cross-talk is charge diffusion.}
\label{psf}
\end{figure}

A particle event can induce cross-talk between neighboring pixels.
This effect increases the number of pixels affected by a particle
impact beyond those directly hit---a potentially significant source of
data loss.  In a detector array, such as those being described here,
cross-talk can occur either in the multiplexer or in the array itself.
In the InSb arrays studied here, pixels are read out every 4
columns---so multiplexer-based cross-talk (``MUX bleed'') will result
in spurious signal 4 columns away from a pixel impacted by a cosmic
ray.  If, for example, one has just read out the pixel at coordinate
$x$=144, $y$=152, the next pixel to be read (in time) will be $x$=148,
$y$=152. Figure~\ref{psf} clearly shows this effect.  The most
important source of cross-talk in the detector array (as opposed to
the multiplexer) is charge diffusion.  Once charges are created in the
photo-conductive layer, their motion is governed by charge diffusion
and the potential established by pixel electrodes.  Holloway (1986)
has numerically solved the charge-diffusion equation for a
2-dimensional detector array.  He found that cross-talk diminished
approximately exponentially with radius.  Moreover, the amount of
cross-talk depends on how far from the electrodes charge is created.
For light, long wavelengths tend to be absorbed deep in the
photo-conductive layer, near the depletion region, and have less cross
talk than shorter wavelengths which would be absorbed at shallower
depths.  Because particle events liberate charge all along their path,
we expect the amount of cross-talk would be intermediate between that
generated by short and long wavelength light.\footnote{Based on
Holloway's (1986) analysis, manufacturers can alter an array's
cross-talk by varying three parameters. These are: (1) the thickness
of the photo-conductive layer, (2) the pixel spacing, and (3) the
diffusion length in the photo-conductor. However, altering any of
these parameters may alter other array properties in undesirable
ways. To cite two examples, an extremely thin detector would have
very little cross-talk at the expense of poor red sensitivity.
Alternatively, one might strive to reduce cross-talk by making big
pixels. However, this would increase pixel capacitance, $C$, and
because the voltage change produced by one charge, d$V$, scales as
${\rm d}V = {\rm d}Q/C$, the resulting array would have
reduced output gain and consequently higher readout noise.}

\begin{figure}
\plotone{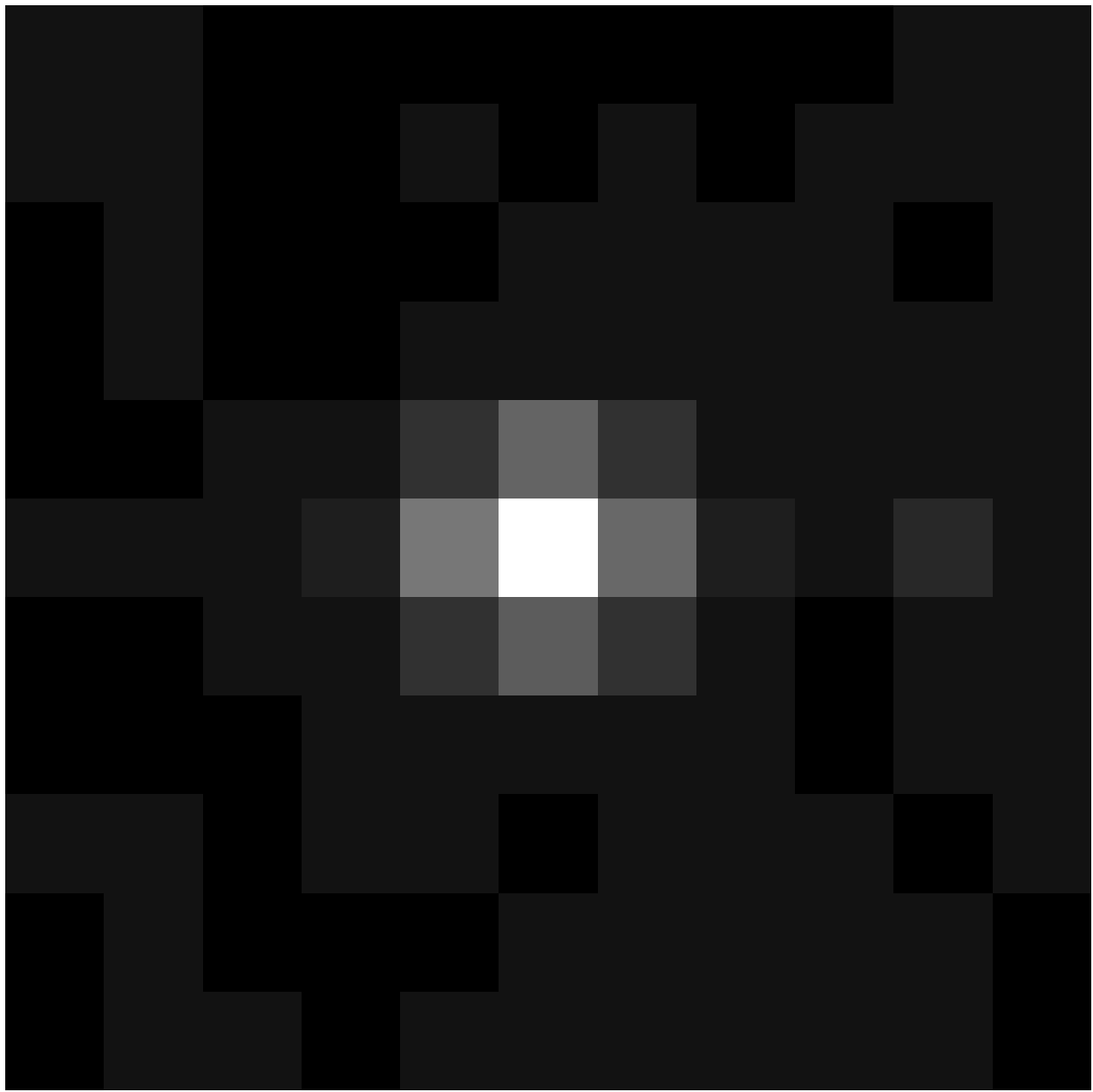}
\caption{Stacking a large number, $\sim$36, of proton hits from the
Si:As data set reveals cross-talk both in the multiplexer and in the
detector array.  This signal from the cosmic ray spread to its
neighbors.  One well-understood mechanism for this cross-talk is
charge diffusion.  The hot pixel to the right of the primary hit is
induced by multiplexer cross-talk.}
\label{psf_SiAs}
\end{figure}

The cross-talk effect was studied and quantified in the InSb data set
for a different study (Rauscher \etal, 2000).  To measure this effect,
Rauscher \etal\ stacked a large number, $\sim$100, of proton hits and
the pixels surrounding them.  We define the term ``edge-neighbors of a
pixel'' to refer to the four pixels which share an edge with the pixel
in question; the ``corner-neighbors of a pixel'' are those which have
a common corner but no common edge with the pixel in question.  In the
InSb data set, the edge-neighbors of a pixel impacted by a particle
hit exhibited a cross-talk-induced signal of about 1.7\% of the signal
in the central, impacted pixel.  The corner-neighbors also showed a
cross-talk-induced signal, about an order of magnitude less than the
edge-neighbors.  Figure~\ref{psf} shows that multiplexer-induced
cross-talk is present in this data set as well, at about the same
magnitude as the corner-neighbors.  The corner-neighbors and the
multiplexer-induced cross-talk are faint with respect to the read
noise, typically at or below $3\sigma_r$, and thus will be
indistinguishable from random noise.  This approach was duplicated for
the Si:As data set with a set of $\sim$36 proton hits.  In the Si:As
data set, Figure~\ref{psf_SiAs}, the magnitude of the cross-talk in
the edge-neighbors was about 3.0\% of the signal in the impacted
pixel, while the corner-neighbors were impacted by about an order of
magnitude less than the edge-neighbors.  There are also signs of
multiplexer-induced cross-talk, also on par with the corner-neighbors.
Again, both the corner-neighbor and multiplexer-induced cross-talk are
below $3\sigma_r$ and should be indistinguishable from random noise.

The net effect of this cross-talk is that most particle events will
affect a clump of pixels in a cross or an extended cross in the case
of a particle which passes through multiple pixels.  In the case of
``glancing'' cosmic rays, which inject relatively small signals, the
signal induced by cross-talk in some or all of the edge-neighbors
could fade into the noise, in which case a smaller clump or only one
pixel might be affected measurably.  One way to increase the ability
of the cosmic ray detection algorithm is to add an ``impugn
neighbors'' step, by which we mean either lowering the threshold for
finding glitches, or rejecting outright, pixels which neighbor a glitch
when one is found.  Such a step will add very little to the running
time, as the Up-the-Ramp software is already dominated by input
(Fixsen \etal, 2000).  However, we are concerned that adding such a
step might cause the cosmic ray identification algorithm to discard
excessive amounts of valid data, to the detriment of the overall data
quality.  Rather than modify the Fixsen \etal\ algorithm, we apply the
algorithm as originally presented to determine the level of its
success without adding an ``impugn neighbors'' step.

The result of a cosmic ray event is the loss of one interval during the
Up-the-Ramp observation (potentially plus a few extra intervals if the
detector shows persistent effects) for a group of pixels clumped in a
(sometimes incomplete) cross-shaped pattern.  This loss degrades the
quality of the data-fitting routine; on average, the signal-to-noise
ratio for a cosmic-ray pixel is reduced by 1/$\sqrt{2}$ for
low-signal cases, up to $\sqrt{(N-1)/N}$ for high-signal cases with $N$
samples.  Of course, this reduction in signal-to-noise will often be
preferable to the alternative, which is total loss.  Furthermore, as
Fixsen, \etal\ point out, on-the-fly cosmic ray rejection enables
longer integration times, which provides a straightforward way to
increase the signal-to-noise to compensate for such loss.

\section{InSb Data Set}\label{sec:InSbData}

The InSb data are from a $256\times256$-pixel InSb (Indium Antimonide)
detector array which researchers at the University of Rochester subjected to
proton flux at the Harvard Cyclotron.  

Each 172-millisecond integration was recorded as one Fowler pair
(Fowler \& Gatley, 1990), followed by a reset of the detector.  As a
result, each raw sample of the detector represents an independent
172 ms integration.  We use a set of 99 such images.  So we can apply the
Up-the-Ramp algorithm to this data set, we co-add the independent
samples to create an Up-the-Ramp sequence (i.e. The Up-the-Ramp
samples are generated from the Raw samples via $U_0~=~R_0$,
$U_i~=~U_{i-1}~+~R_i$).  The set $\{U_{0...98}\}$ approximates a
uniformly-sampled data set; for example, the covariance of the samples
does not match the form of the covariance assumed in the data
processing algorithm (Fixsen, \etal, 2000).  Since this is an
approximation to a Up-the-Ramp sequence, we are making use of
less-than-ideal data to conduct this test.  However, the differences
should be slight and should not affect the overall validity of the
results of this test.

The array was irradiated using a 70 MeV beam; the dewar walls
attenuated the energy to 20 MeV.  The Rochester team conducted these
tests with the proton beam emitting
$6.5~\times~10^4$~protons~cm$^{-2}$~s$^{-1}$.
From the array pixel size of $30\mm~\times~30\mm$ pixels and
the fact that there are $256~\times~256$ pixels in the detector, we
find that the array is impacted by approximately
$3.8\times10^4$~protons~sec$^{-1}$.  This is approximately equal to
the ion flux which would be expected in deep space at the height of a
major solar event, between 3 and 4 orders of magnitude greater than the
``typical'' cosmic ray flux in deep space, and so represents a severe
test of the cosmic ray detection algorithm. (Barth \& Isaacs, 1999;
Tribble 1995).

The path through the detector array of a cosmic ray from an isotropic
distribution is effectively random.  However, for the InSb radiation
test, the proton beam had a predetermined angle of incidence, 23 degrees
from the plane of the detector.  Coupled with the InSb pixel
dimensions of $30\mm~\times~30\mm~\times~8\mm$, we calculate
that the projection of the typical proton's path onto the plane of the
detector array is about 19$\mm$ long.  If we project this path length onto
random locations on the plane of the detector, we find that about half
of the protons will pass through 1 pixel and about half will pass
through 2 pixels.  For this computation, we ignore the effects of
scattering and secondary particles.

When a proton physically passes through 1 pixel, cross-talk effects
will cause a total of 5 pixels to be affected at the $3\sigma$ level
or higher.  When a proton passes through 2 pixels, 8 total pixels are
affected at $3\sigma$ or higher.  If we assume a 50-50 split between
these two cases for the InSb data set, an average of 6.5 pixels are
affected per proton. Since we expect the InSb detector array will
encounter $3.8\times10^4$~protons~sec$^{-1}$, we find that about
13.8\% of the detector array is affected by radiation for each 172 ms
second sample.  This type of analysis presents difficulties in
performing formal uncertainty analysis, but we estimate that these
numbers are good to $\pm25$\%.

\begin{figure}
\plotone{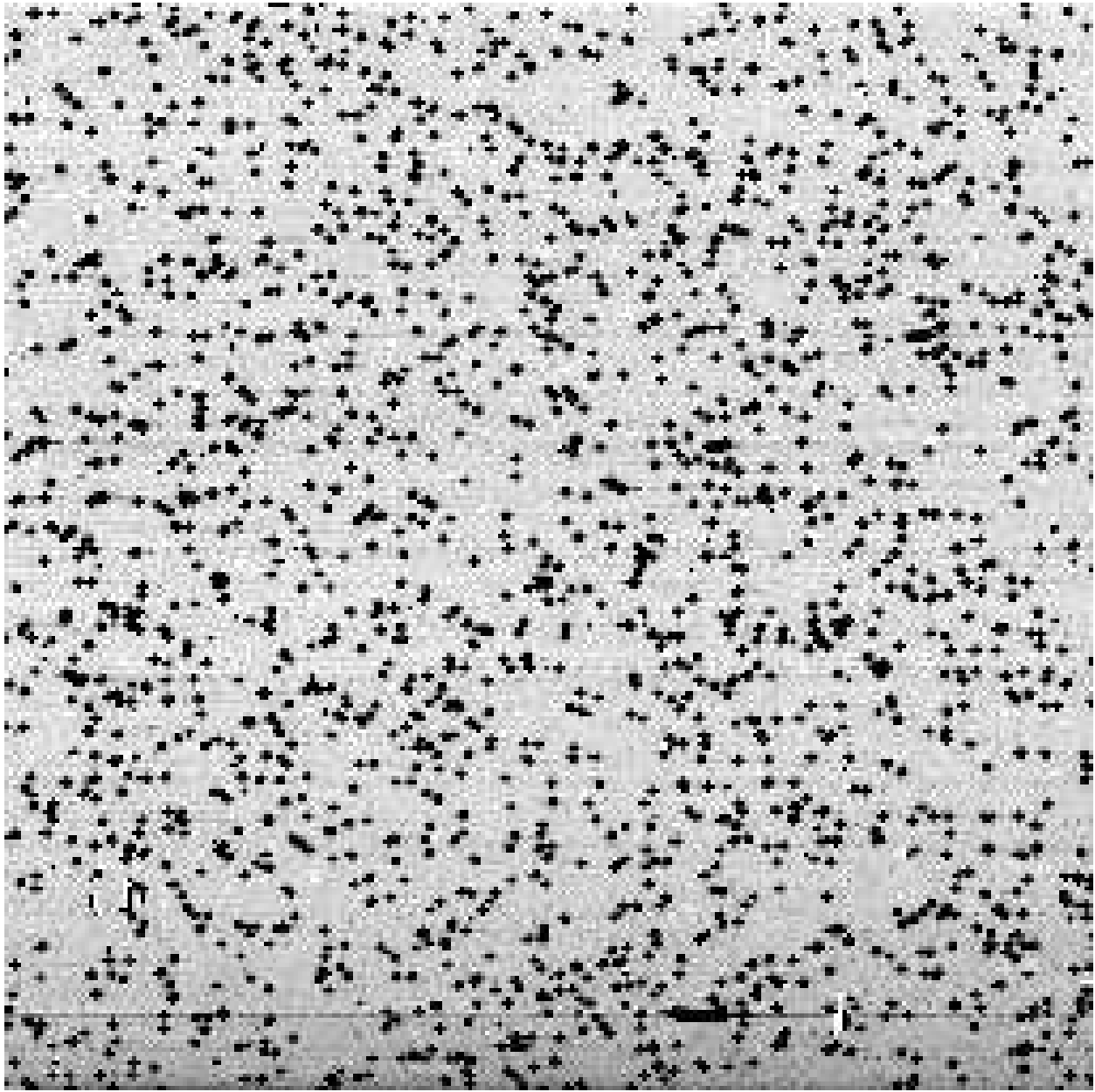}
\caption{An Up-the-Ramp sample from the InSb data sequence.  This is a
raw sample ($R_{74}$ in the notation used in Section~\ref{sec:InSbData}),
chosen at random from the observation sequence.  Dark pixels in this
figure correspond to high-signal regions, light pixels correspond to
low-signal regions.  Most of the high-signal regions of the detector
are due to proton impacts (compare to Figure~\ref{InSbMedian} to see
the detector ``hotspots'').  Since this is the raw difference image,
the number of particle hits on the detector is the amount that
accumulate in 172 milliseconds.}
\label{InSbRaw}
\end{figure}

A sample image from the raw InSb data set appears in
Figure~\ref{InSbRaw}.  This is a raw image, randomly chosen from the
original sequence ($R_{74}$ in the notation used earlier), and the
number of cosmic rays on this image is that of a single 172
millisecond interval.  The images are mostly dark frames.  To create a
``pseudo-real sky'' image for comparison purposes, we create a median
image, Figure~\ref{InSbMedian}, in which each pixel is the median
value of the raw samples for that pixel.  We then scale to the full
integration time by multiplying the median values by 99, the number of
samples in the observation sequence.  The result compares favorably
with an illuminated flat field frame.  The observed ``tree ring''
structure, representing doping density variations in the InSb, appears
in illuminated frames but not in dark current frames.  Thus, either
there is a small light leak, or the filter wheel has been warmed by
proton irradiation, illuminating the array, albeit to a very low
level.  Similarly, there is a readout ``glow''; this effect is
present over the entire image but is most pronounced in the first few
rows sampled (corresponding to the bottom of the frame).

\section{Si:As Data Set}\label{sec:SiAsData}

The Si:As data are from a $40\times40$-pixel region of a
$256\times256$-pixel Si:As (Silicon doped with Arsenic) IBC detector which
was subjected to proton flux at the University of California--Davis
cyclotron. 

The Si:As data set is a series of 52-millisecond integrations, each
recorded as a 1 Fowler-pair observation (Fowler \& Gatley, 1990),
followed by a reset of the detector.  A $40\times40$-pixel subregion
of the detector was used; the photo-conducting region of each pixel
was approximately $30\mm~\times~30\mm~\times~30\mm$.  As in the InSb
case, the data are dark frames.

We generate an approximate Up-the-Ramp sequence by co-adding 36
independent samples using the same procedure described in
Section~\ref{sec:InSbData}.  The caveats regarding the use of this
approximation which were discussed in Section~\ref{sec:InSbData} apply
to the Si:As data as well.  These differences still should not
affect the overall validity of the results of this test.

The array was irradiated with a 67 MeV proton beam operating in an
uncalibrated mode (i.e. the proton flux is not independently
known).  The proton beam passed through only a thin plastic window
on its way to the detector---there was no dewar wall in the
proton beam path, as there was in the InSb data set.  As a result, the
protons should not be scattered or attenuated appreciably.  

As the proton flux was not directly measured, the only source of
information with regard to the number of events expected on the
detector is the data set itself.  We estimate the number of events by
examining a sequence of 36 raw (difference) samples and counting the
number of outliers from the background dark value as a function of the
outlier threshold; we expect the read noise to be a Gaussian
distribution and thus the number of outliers due to read noise should
fall off in a Gaussian pattern as we increase the threshold.
Outliers caused by protons, which should be large with respect to the
dark value, will not fall off at low multiples of $\sigma$.  In the
sample we tested, the number of outliers flattened out in the interval
4$\sigma$-to-5$\sigma$.  If we use these values as the lower and upper
limits for the threshold, we find that between 3.0\% ($4\sigma$) and
2.4\% ($5\sigma$) of the pixels should be identified as particle hits
or neighbors impacted by cross-talk for each interval.  For a 36-frame
sequence of $40\times40$-pixel images, we expect to find around
$1500~\pm~200$ pixels impacted by cosmic rays during the observation.
This figure includes pixels impacted both by cosmic rays and by
cross-talk (see Section~\ref{sec:interaction}).  Since the proton beam
was oriented normal to the plane of the detector, most of the protons
would physically pass through one pixel; based on the cross-talk
model, we expect that $\sim$20\% of the glitches are cosmic-ray hits,
the rest being caused by cross-talk.  We find that the cosmic ray flux
is approximately $1.1\times10^4$~protons~s$^{-1}$~cm$^{-2}$.  This
flux is about a factor of 6 lower than that used in the InSb test.

The signal in the Si:As images is heavily quantized (i.e. the
histogram of the data values is fairly sparse).  The statistical tests
used to identify the cosmic ray glitches and the line-fitting
algorithm from Fixsen, \etal\ will perform best when the data are
continuous or the level of quantization is small relative to the size
of the data.  Although we do not expect this to be a major problem,
the uncertainty in the results from the line-fitting routine may be
larger than it would otherwise be.

\begin{figure}
\plotone{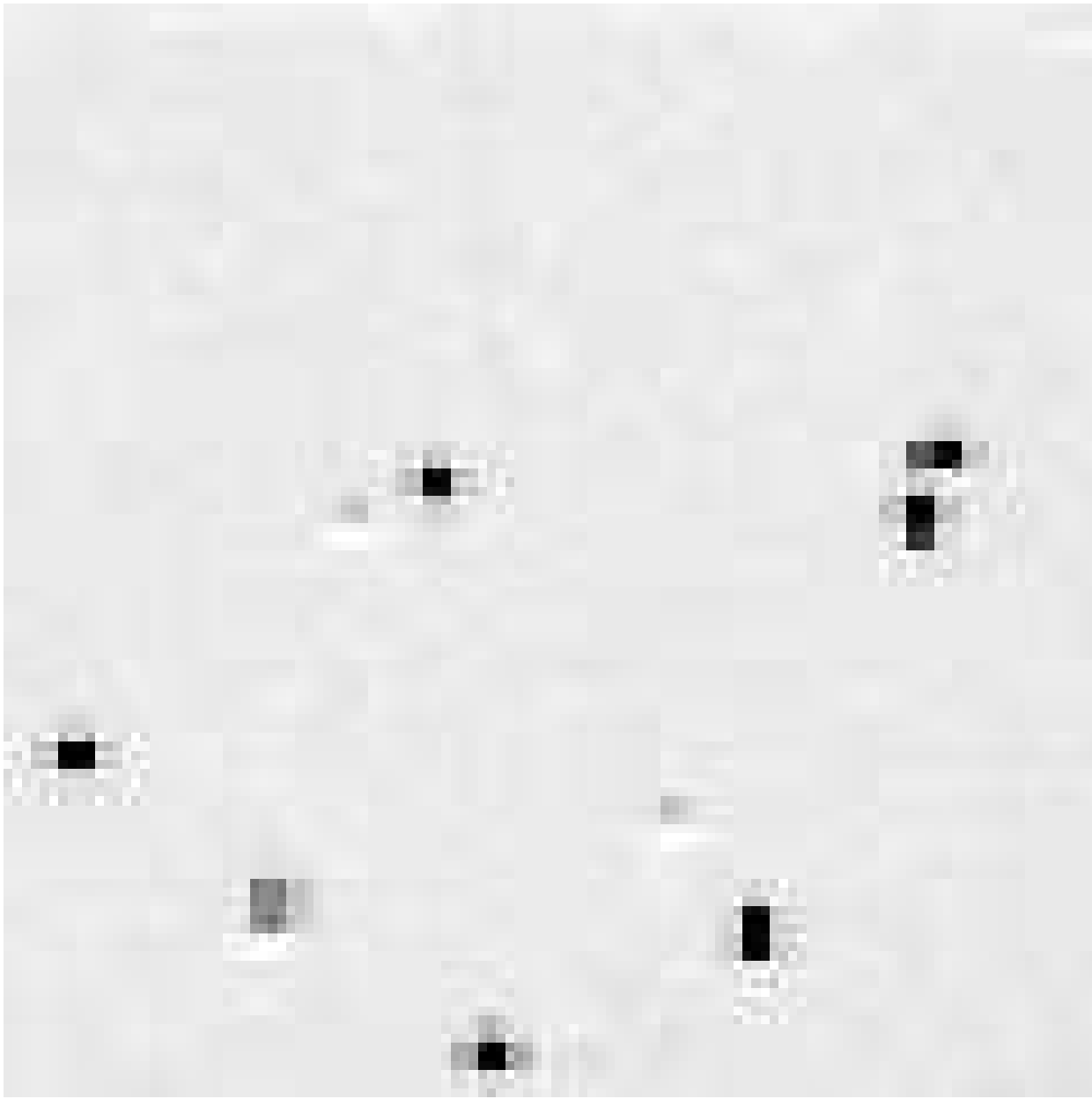}
\caption{An Up-the-Ramp sample from the Si:As data sequence.  This is
a raw sample ($R_{27}$ in the notation used in
Section~\ref{sec:InSbData}), chosen at random from the observation
sequence.  Dark pixels in this figure correspond to high-signal
regions, light pixels correspond to low-signal regions.  Most of the
high-signal regions of the detector are due to proton impacts (compare
to Figure~\ref{SiAsMedian}).  Since this is the raw difference image,
the number of particle hits on the detector is the amount that
accumulate in 52 milliseconds.}
\label{SiAsRaw}
\end{figure}

A sample image from the raw Si:As data set appears in
Figure~\ref{SiAsRaw}.  This is a raw image, randomly chosen from the
original sequence ($R_{27}$ in the notation used in
Section~\ref{sec:InSbData}), and the number of cosmic rays on this image
is that of a single interval, 52 milliseconds.  To create a
``pseudo-real sky'' image for comparison purposes, we create a median
image, Figure~\ref{SiAsMedian}, in which each pixel is the median
value of the raw samples for that pixel.  We then scale to the full
integration time by multiplying the median values by 36, the number of
samples in the observation.

\section{InSb Results \& Discussion}\label{sec:InSbDiscussion}

\begin{figure}
\plotone{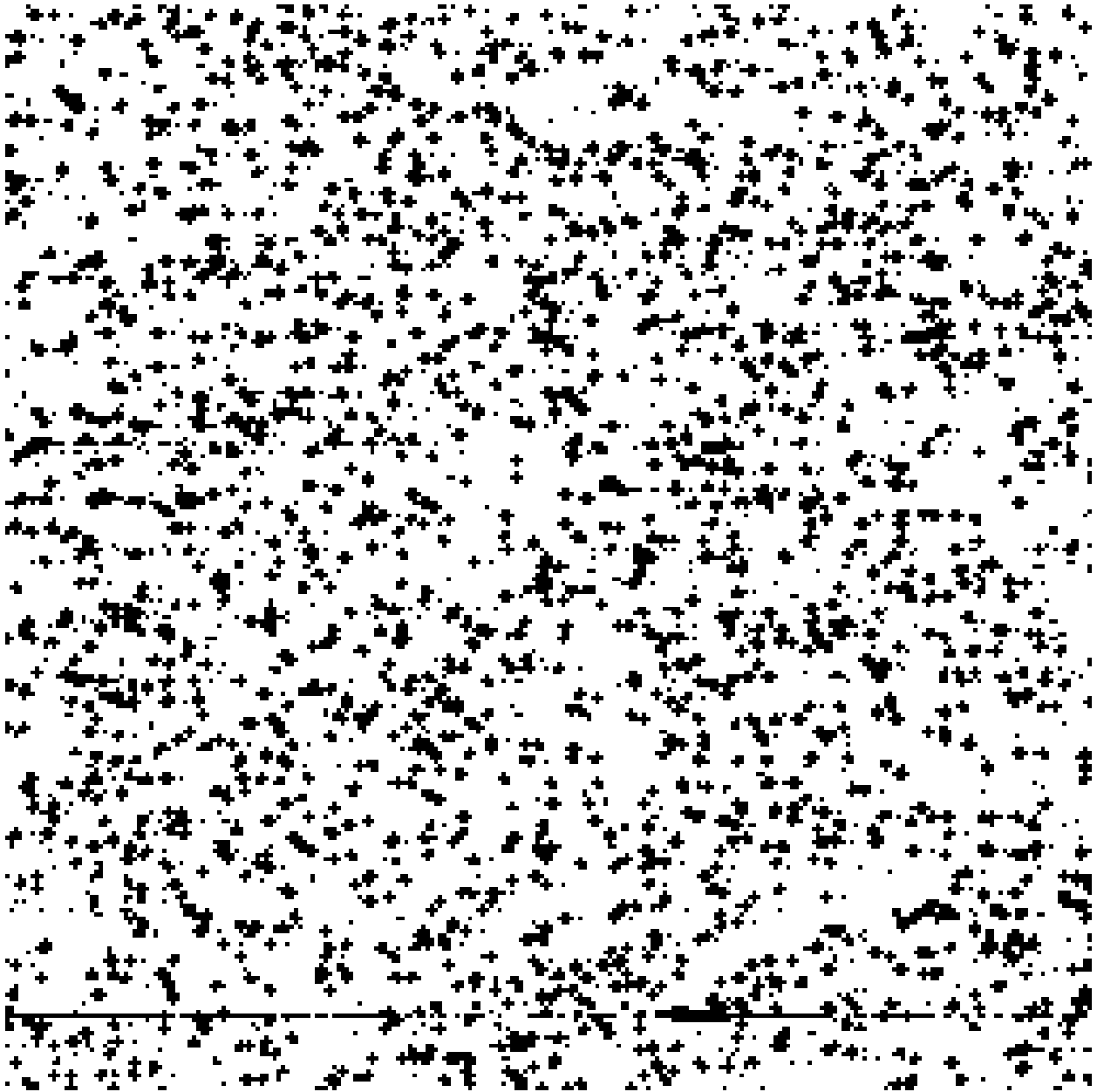}
\caption{The cosmic ray mask for the sample shown in
Figure~\ref{InSbRaw}.  The dark pixels show the places where cosmic
rays were identified and removed from this sample.  Note that the
cross-hatching pattern matches the cross-talk pattern described in
Section~\ref{sec:interaction}.}
\label{InSbMask}
\end{figure}

The output data image from the processing the 99 InSb samples appears in
Figure~\ref{InSbResult}.  A difference image between this image and
the median image appears in Figure~\ref{InSbDifference}.
Figure~\ref{InSbMask} shows the pattern of cosmic rays identified in a
randomly-chosen sample (the same sample that is shown in
Figure~\ref{InSbRaw}).

\begin{figure}
\plotone{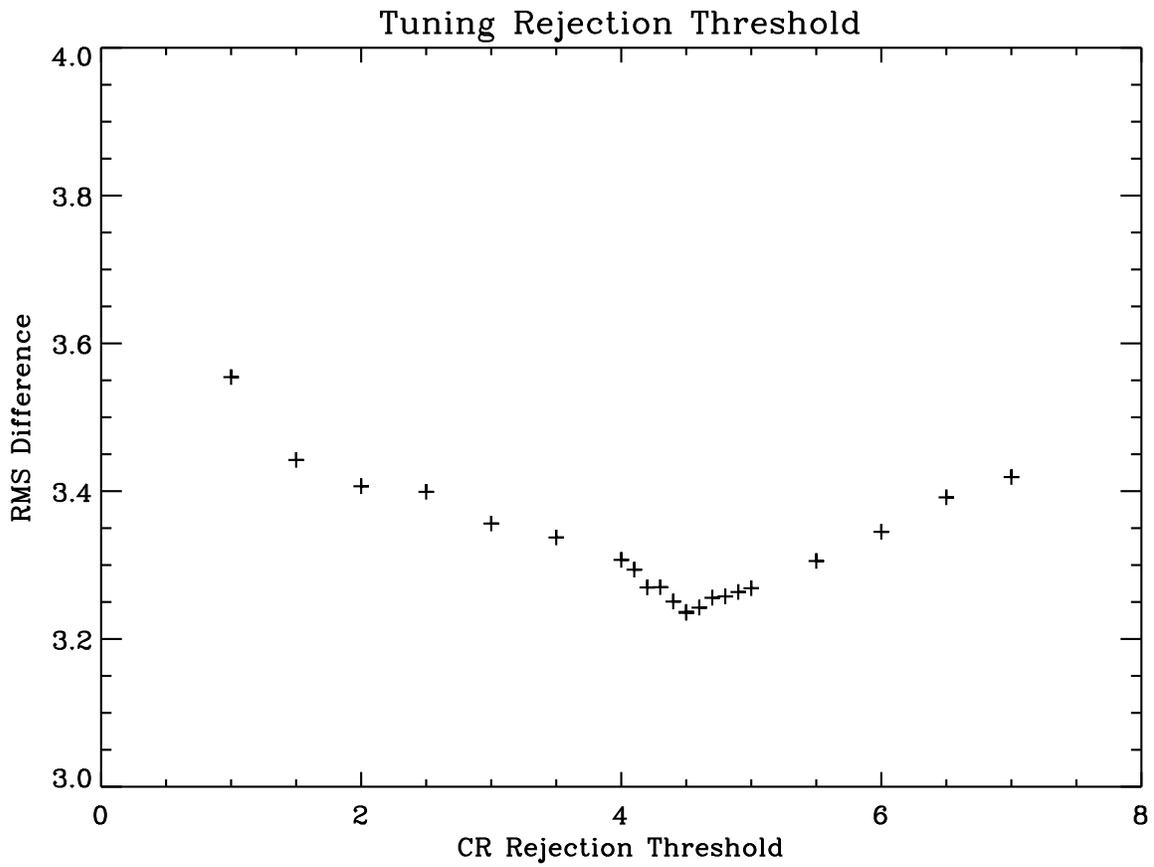}
\caption{The RMS differences for the InSb data set plotted as a
function of cosmic ray rejection threshold.  The minimum occurs around
$4.5\sigma$.  The value of $4.5$ is used for the cosmic ray rejection
threshold for all results discussed in this paper.}
\label{InSbRMS}
\end{figure}

We perform one simple tuning operation with this data set to verify
that the cosmic ray threshhold of $4.5\sigma$ used by Fixsen, \etal\ 
was appropriate for this data set.  We divide the 99-image sequence of
InSb samples into two 49-image sequences (discarding one), and
generate two Up-the-Ramp sequences.  The two sequences are then
processed using a series of cosmic ray rejection threshholds,
generating one pair of output images for each threshhold value.  We
measure the uncertainty of the threshhold value as the RMS difference
between the two images of each pair for that value.  We examine the
threshholds to be in the range of $1-7\sigma$ equally spaced at
intervals of $0.5$ (i.e. $[1,1.5,2,2.5,...,7]\sigma$).  The
initial pass shows a minimum between $4\sigma$ and $5\sigma$, so we
repeat the test between these two values at intervals of $0.1$.  The
results are shown in Figure~\ref{InSbRMS}; the minimum appears at
$4.5\sigma$.  Because there are insufficient samples to repeat this
test reliably with the Si:As data set, we adopt $4.5\sigma$ as the
threshhold value for that data set as well.  We also find median image
for the two subimage sequences, and compute the RMS difference for the
two median images.  The RMS difference at $4.5\sigma$ is 3.24,
compared with the value of 2.33 for the median case; however, if we
discard the pixels for which contain a surviving cosmic ray in either
image of each set (27 pixels in the Up-the-Ramp case, 49 pixels in the
median case), the Up-the-Ramp RMS difference drops to 1.34, whereas
the RMS for the median images is 1.37.  

As discussed in Section~\ref{sec:InSbData}, we find that 13.8\% of the
detector should be affected by a proton or by cross-talk for each
sample; the total number of pixels affected by protons plus those
affected by cross-talk comes to 895,000 for the entire sequence ($\pm
224,000$, using our ballpark estimate of 25\% for the uncertainty in
this number).  The cosmic ray rejection algorithm, with the minimal
tuning, identifies and rejects 1,063,000 bad samples from this
observation sequence.  This number lies within the uncertainty range
for the number of expected bad samples.  We must be cautious against
reading too much into this result, but it is safe to say that the
number of cosmic rays identified is in the right
ballpark\footnote{Those who dislike sports analogies can substitute
``...is enough egg for an omlette.''}.  The overall performance can
be improved by tuning the parameters of the algorithm, which will
require repeated and detailed radiation tests with the detector.

An examination of Figure~\ref{InSbMask} shows that most of the cosmic
ray detections are clumped in cross-hatch patterns, matching the
current-leak pattern described in Section~\ref{sec:interaction}.
Thus, the cosmic-ray identification algorithm catches many of the
neighboring pixels affected by cross-talk without having to add an
``impugn neighbors'' step, as described in Section~\ref{sec:interaction}.
When the full energy spectrum of cosmic rays in deep space is
considered, the amount of charge injected, and the resulting signal,
will often be smaller than that induced by a cyclotron proton.  The
fact that the cosmic ray identification algorithm catches the
neighbor-leak events shows that it is capable of identifying the full
spectrum (i.e. bright, faint and in-between) of cosmic rays that we
expect to encounter in a deep space environment (contrasted with the
limited particle energy spectrum generated by the cyclotron). As this
is a dark frame, faint bad pixels stand out from their neighbors; the
current cosmic ray identification software might not work as well in
brighter regions of the image, where increased Poisson noise will
affect the statistical test used to identify cosmic rays.  Brighter
images and fainter cosmic rays will combine to mask the
``neighbor-leak'' events---tuning the algorithm to the specific
detector would minimize the impact of these effects, but a step to
``impugn neighbors'' might still be required during astronomical
observations.

\begin{figure}
\plotone{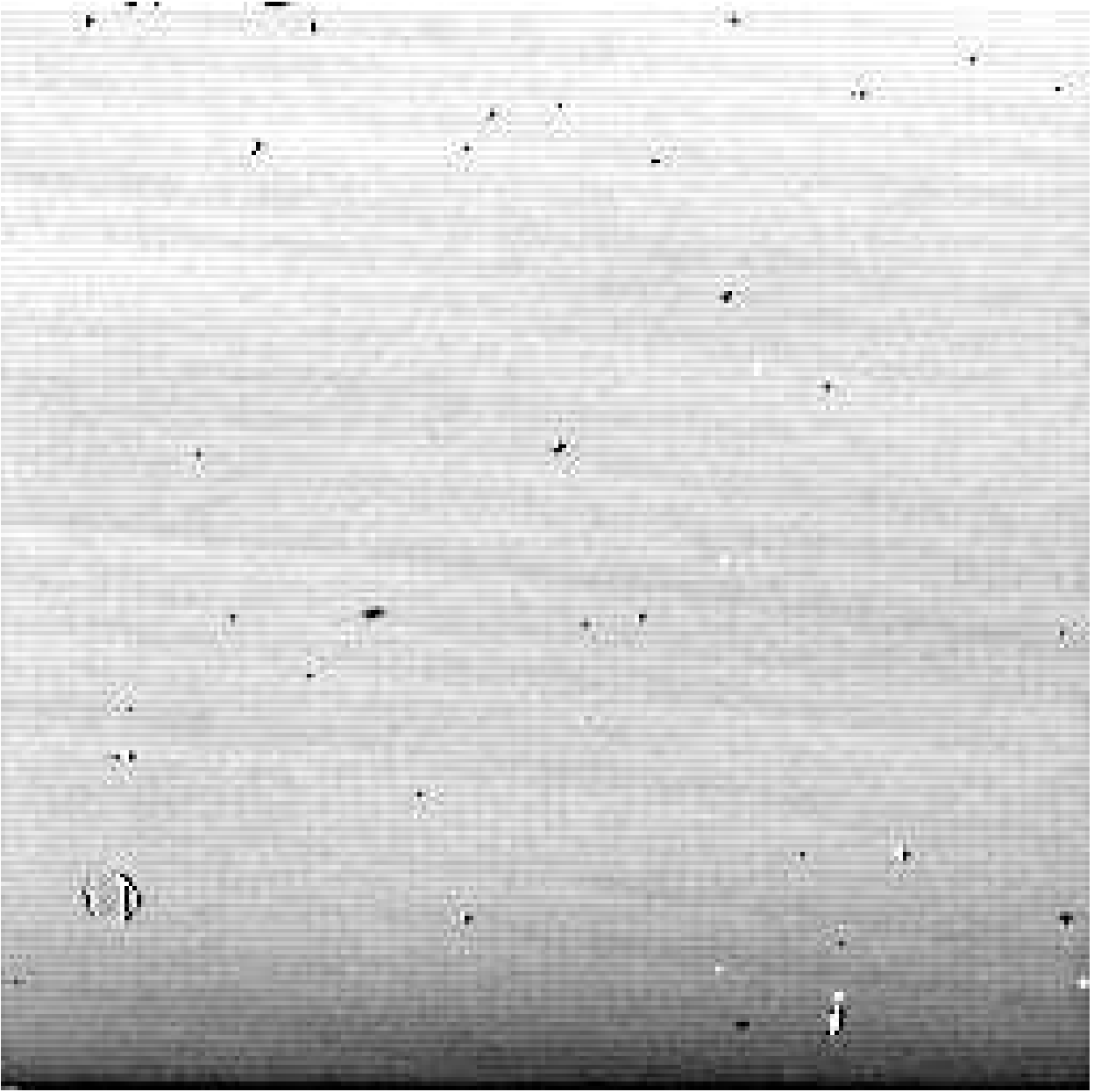}
\caption{The median image generated from the raw InSb samples.  This
image was generated for comparison purposes: each pixel is the median
value at that pixel across the 99 samples which constitute the raw
data set.  Most of the high-signal regions are due to ``hot spots'' on
the detector.  Also, note the readout glow across the bottom of the
detector and the ``tree ring'' structure across the image.}
\label{InSbMedian}
\end{figure}

\begin{figure}
\plotone{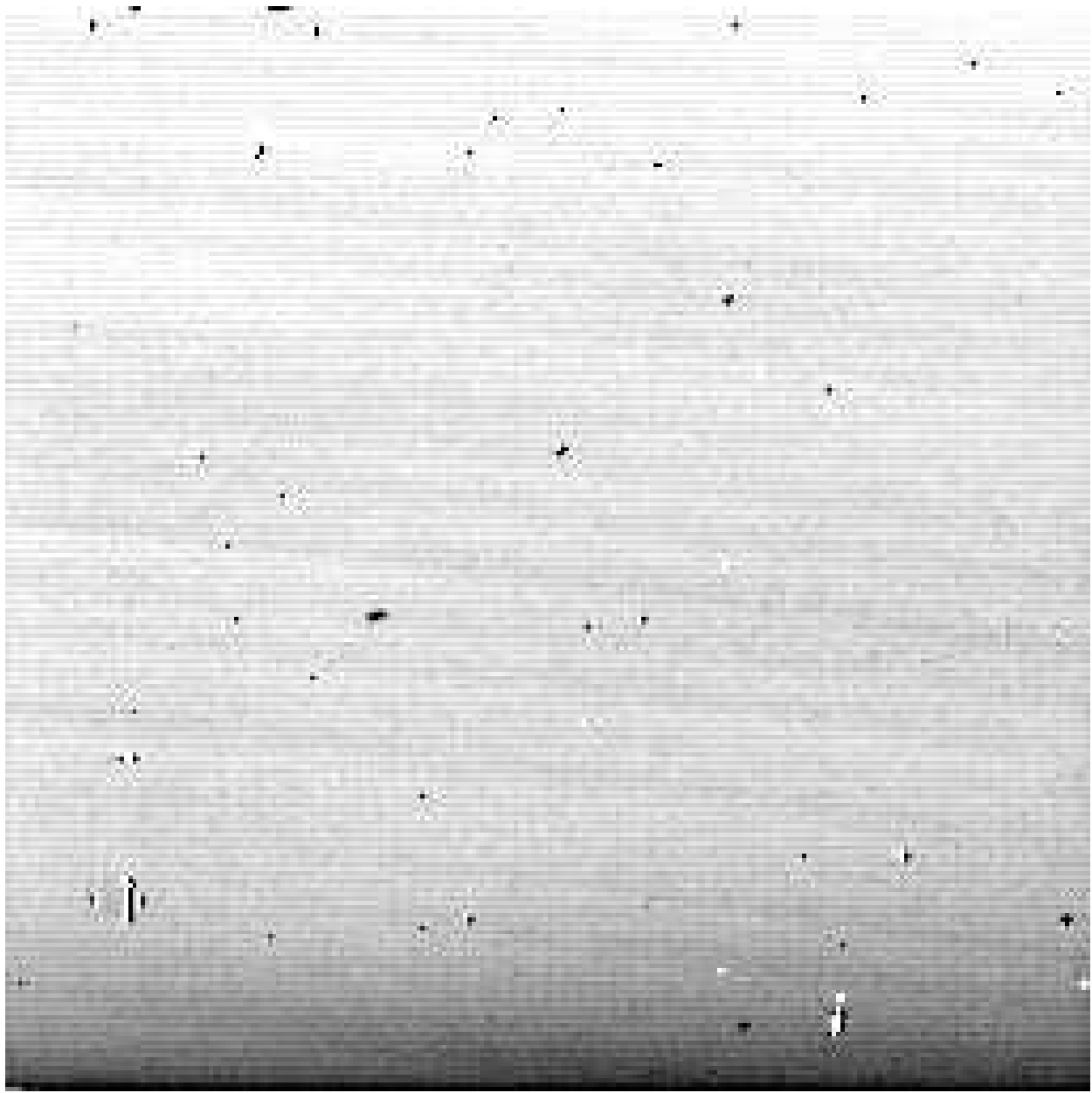}
\caption{The output data image from processing 99 InSb data samples.
Compare with Figure~\ref{InSbMedian}---there is a high correspondence
between high-signal pixels in this image and those in
Figure~\ref{InSbMedian}.  Note that the glow at the bottom of the
detector and the ``tree ring'' structure are preserved in this output
image.}
\label{InSbResult}
\end{figure}

\begin{figure}
\plotone{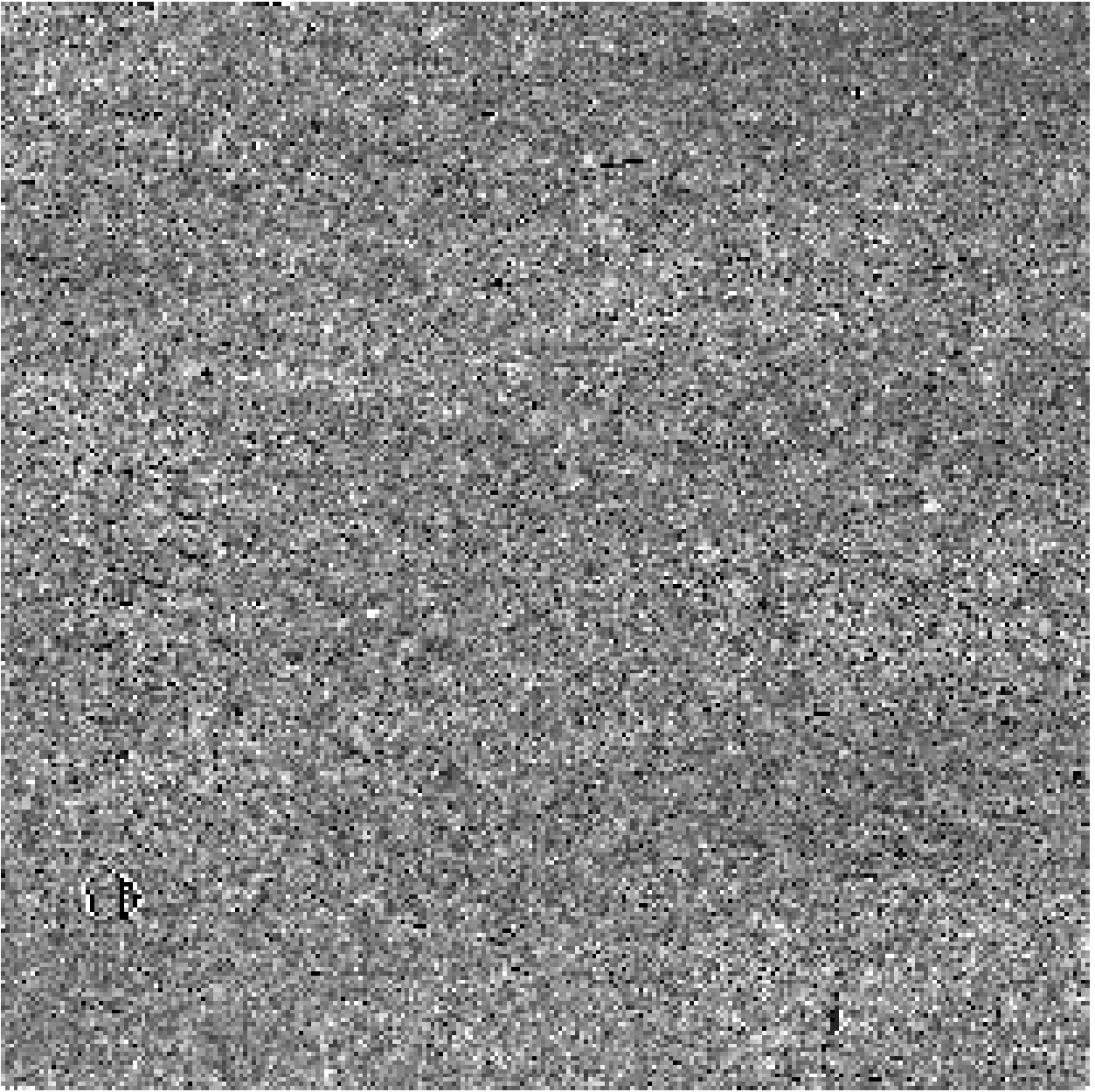}
\caption{The difference between Figure~\ref{InSbResult} and
Figure~\ref{InSbMedian}.  In this image, dark pixels correspond to
regions where Figure~\ref{InSbResult} has a higher signal; light pixels
correspond to regions where Figure~\ref{InSbMedian} has a higher
signal; regions where the two images are equal are grey.  The greyscale spread covers $\pm3\sigma_{r}$.}
\label{InSbDifference}
\end{figure}

The frames in this study are dark frames.  As was discussed in
Section~\ref{sec:InSbData}, there appears to be a small light leak or
thermal glow which is faintly illuminating the array as a whole.
There is also an apparent ``glow'' effect, which is most prominent in
the first few rows sampled (the bottom of the image in the orientation
shown in Figure~\ref{InSbMedian} and Figure~\ref{InSbResult}).  The
only other non-random contributions to the flux should be the dark
current, ``hot'' pixels and other detector artifacts.  By taking the
image generated from performing the median operation on the raw data
set, Figure~\ref{InSbMedian}, and scaling up from 1 sample to 99 by
multiplication, the dark value is computed to be $4880~\pm~41$ data
units (1.8 electrons per data unit).  The dark value on the processed
image is computed to be $4881~\pm~40$ data units---the median and
processed median dark values agree very strongly, well within the
margin for error. In addition, note that the readout ``glow'' at the
bottom of the image as well as the faint ``tree ring'' structure over
much of the detector, both of which appear in Figure~\ref{InSbMedian},
are preserved in Figure~\ref{InSbResult} (both of those effects can be
mistaken for errors in printing or reproduction, but they are
genuine).

Overall, the processed image and the median image agree very well at
most individual pixels to within the uncertainty of the dark level
established in the previous paragraph.  99.5\% of all
pixels agree within this tolerance; most of the remainder are
surviving cosmic rays.  This result, combined with the estimated
number of cosmic ray events derived earlier, suggests that the cosmic
ray algorithm successfully identified and removed over 99.96\% of the
proton events on the detector.

We examine in detail the pixels in which the processed image and the
median/comparison image disagree by at least $3\sigma$.  A total of 88
pixels fit into this category; 44 of them occur in places where both
of the images are high-signal (detector hotspots), and thus where we
expect that higher Poisson noise, non-linear response and other
effects increase the normally-expected noise.  Of the remaining 44
instances, 35 are cases where the processed image is brighter than the
median image; these are cosmic rays which were missed by the cosmic
ray detection algorithm.  The remaining 9 instances, where the
processed image is darker, are all cases where more than half of the
samples were affected by cosmic rays, so the median of the sample is a
cosmic ray sample---these 9 pixels contain cosmic rays which were
caught by the algorithm in Fixsen, \etal, but which the median
operation missed.

\section{Si:As Results \& Discussion}\label{sec:SiAsDiscussion}

\begin{figure}
\plotone{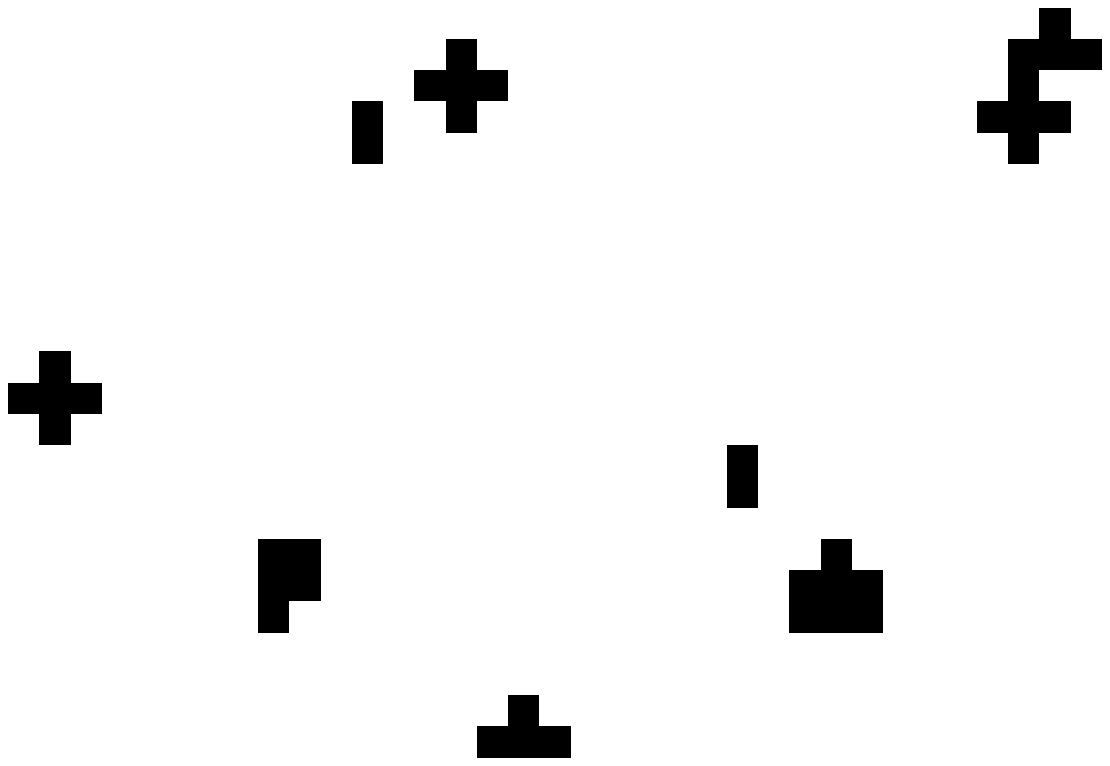}
\caption{The cosmic ray mask for the sample shown in
Figure~\ref{SiAsRaw}.  The dark pixels show the places where cosmic
rays were identified and removed from this sample.  Note that the
pattern of glitches is similar to the cross-hatching pattern described
in Section~\ref{sec:interaction}.}
\label{SiAsMask}
\end{figure}

The output data image from the processing the 36 Si:As samples appears
in Figure~\ref{SiAsResult}.  A difference image between this image and
the median image appears in Figure~\ref{SiAsDifference}.
Figure~\ref{SiAsMask} shows the pattern of cosmic rays identified in a
randomly-chosen sample (the same sample that is shown in
Figure~\ref{SiAsRaw}).

As discussed in Section~\ref{sec:SiAsData}, we expect to find
$1500~\pm~200$ pixels affected by a proton or cross-talk during the
full 36-sample observation.  The cosmic ray rejection algorithm, with
only the minimal tuning described in Section~\ref{sec:intro},
identifies and rejects 1230 bad samples during the observation
sequence.  This number lies outside of the uncertainty range for the
number of expected bad samples, but by a small margin.  Again, the
result is promising.

An examination of Figure~\ref{SiAsMask} shows that most of the cosmic
ray detections are clumped in cross-hatch patterns, again matching the
current-leak pattern described in Section~\ref{sec:interaction}.  As
was the case with the InSb case, this shows that the cosmic-ray
identification algorithm catches the neighbors affected by
cross-talk without requiring any extra steps.  

\begin{figure}
\plotone{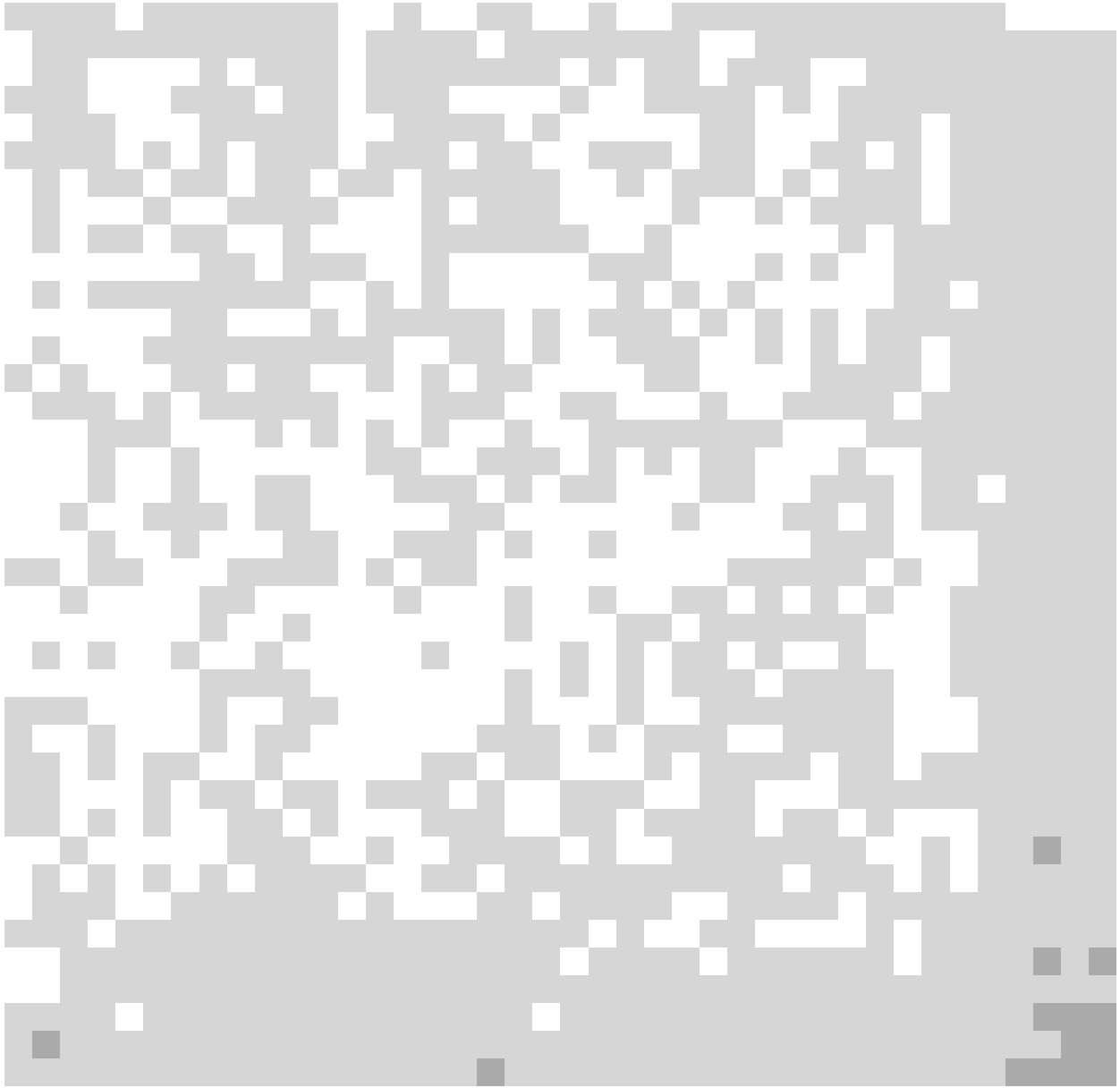}
\caption{The median image generated from the raw SiAs samples.  This
image was generated for comparison purposes: each pixel is the median
value at that pixel across the 36 samples which constitute the raw
data set.  The region sampled for this test is $40\times40$ pixels;
the pixellation in this image and the other Si:As images is
caused by scaling the image up to the size shown here.  The data in the Si:As set are severely quantized, which reduces the quality of this image.}
\label{SiAsMedian}
\end{figure}

\begin{figure}
\plotone{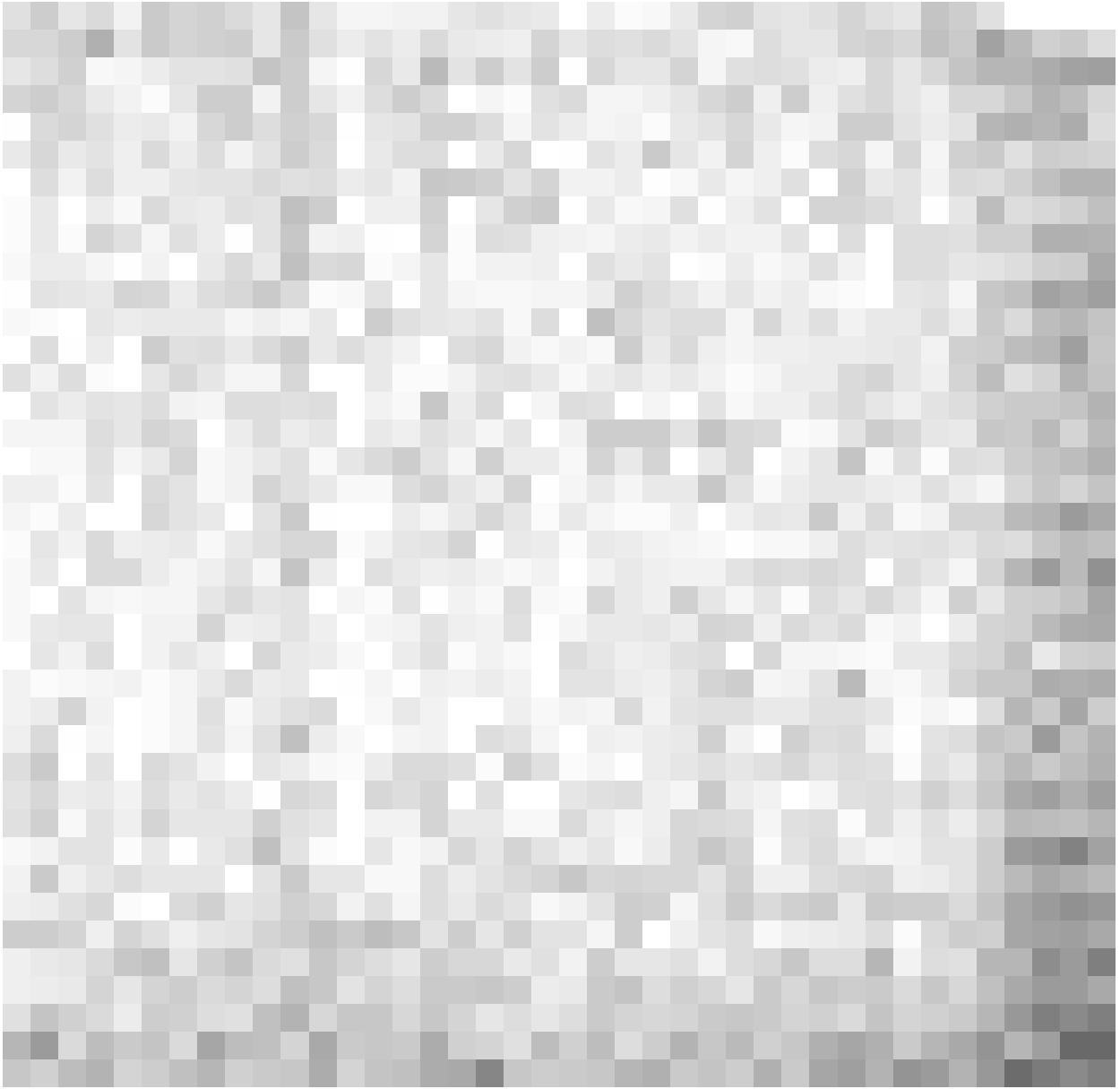}
\caption{The output data image from processing 36 Si:As data samples.
Compare with Figure~\ref{SiAsMedian}.}
\label{SiAsResult}
\end{figure}

\begin{figure}
\plotone{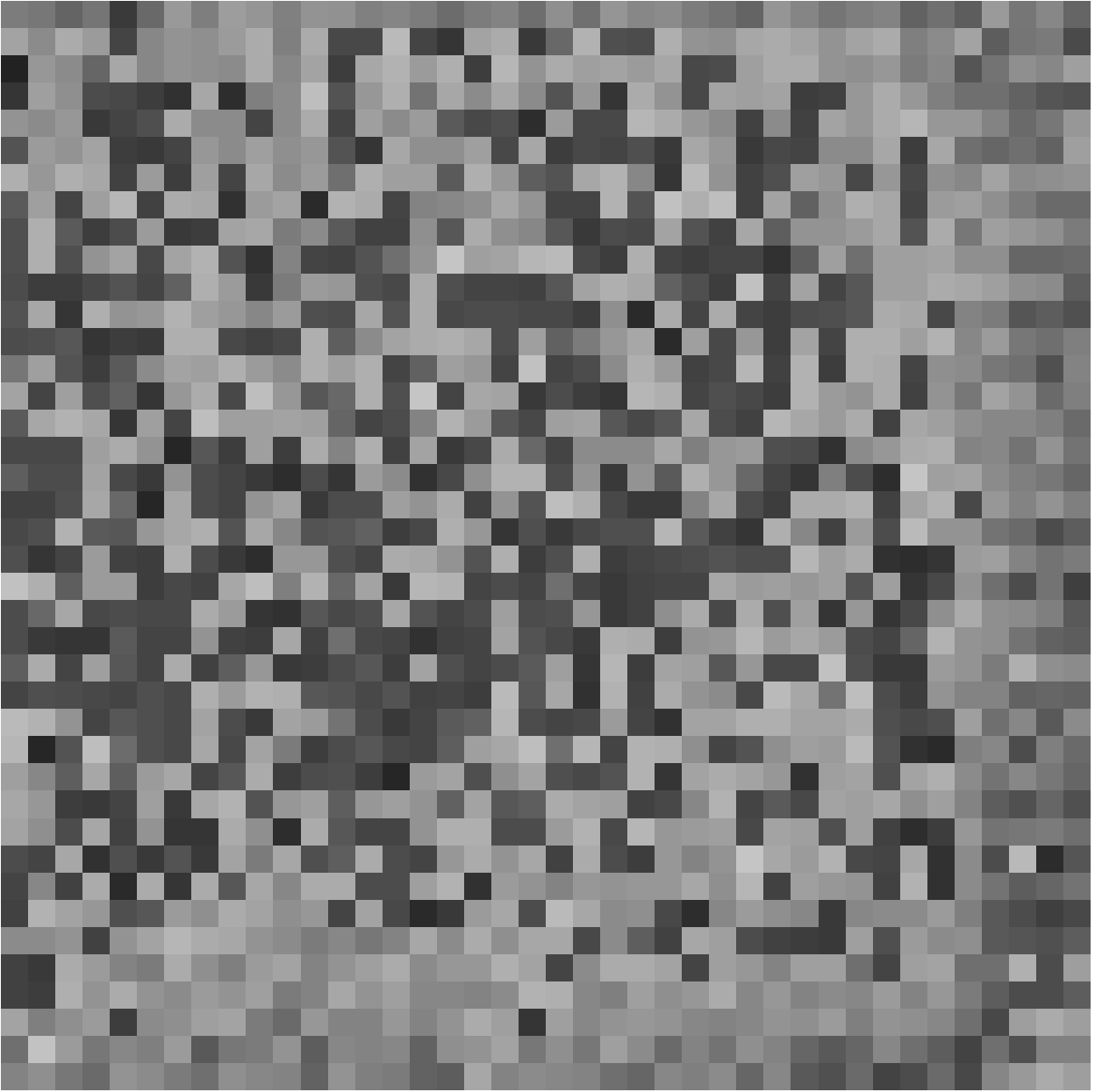}
\caption{The difference between Figure~\ref{SiAsResult} and
Figure~\ref{SiAsMedian}.  In this image, light pixels correspond to
regions where Figure~\ref{SiAsResult} has a higher signal; dark pixels
correspond to regions where Figure~\ref{SiAsMedian} has a higher
signal; regions where the two images are equal are grey.  Overall,
there is no spatial difference or trend.  The quality of this image is
limited by the severe quantization of Figure~\ref{SiAsMedian}.}
\label{SiAsDifference}
\end{figure}

The Si:As images are dark frames; the only non-random signals are from
dark current and detector artifacts.  By taking the median image
(Figure~\ref{SiAsMedian}) and scaling from 1 sample to 36 by
multiplication, the dark value is found to be $375~\pm~9$ data units,
compared to the processed image's dark value of $348~\pm~6$ data
units---the dark values for these images agree to within 3$\sigma$.
If we compare Figure~\ref{SiAsMedian} and Figure~\ref{SiAsResult}, we
see that the signal along the bottom and right-hand side of the
detector are preserved in the output image, although heavy
quantization limits the amount of observable structure in the median
image.

Overall, the processed and median images disagree by more than 3 times
the uncertainty of the processed image at 5 pixels.  In all 5
cases, the processed image has a lower signal than the median image by
just over $3\sigma$ (3.01--3.12$\sigma$).  These deviations appear
to be caused by error in the data fitting routine due to the heavy
quantization of the input data.  As the proton hits will bias the
median image towards higher signals, we recomputed the median image,
discarding reads which differ from the mean by more than $3\sigma$.
Removal of this bias reduces the number of different pixels to 3.  As
the large quantization adds noise to the median which will be less
pronounced in the processed data, the median may be a less accurate
measure of the ``real sky'' than the output image.  There do not
appear to be any missed cosmic rays in the output image.

\section{Conclusion}\label{sec:conclusion}

The Up-the-Ramp sampling and on-the-fly cosmic ray rejection algorithm
performs excellently on the radiation test data from the InSb and
Si:As detectors, particularly when we consider that the data were not
originally Up-the-Ramp samples (but co-added intervals), the algorithm
parameters were only minimally tuned to the detector, the data were
not linearized before processing and we ignored any possibility of
persistent radiation effects, which is particularly important since
the particle rates were extremely high.  Detailed tuning of the
algorithm with the experiment detector will improve these results.
The cosmic ray algorithm did not introduce any 2-D structure into the
image, and preserved the structure which was there.  Finally, the
small number of pixels for which the median and processed images which
differed significantly shows that the Up-the-Ramp sampling
can and does preserve the photometric quality of the data.

We have attempted to answer some common concerns about the use of
Up-the-Ramp sampling strategies.  We have shown that the algorithm
performs well without precise tuning to the detector characteristics,
as we have obtained good performance without such tuning.  Another
concern is that the cosmic-ray/detector interaction model assumed in
the design of the algorithm is too simplistic; however, the results
here show good performance without requiring additional complicating features.
In short, we haven't found any ``show-stoppers'' that would lead to
the conclusion that the algorithm requires excessive computing or
human interaction to provide useful scientific data.

These results are limited to the detector technologies being studied
here.  A key future direction for this study is to apply the
techniques discussed here to additional detector technologies (such as
HgCdTe technology, used in the NICMOS instrument on HST).  This is
particularly true in the context of detector characterization studies
for observatories and instruments being designed currently or in the
near future, which is how this study started.  However, the fact that
we obtain very similar results with the two technologies suggests that
the cosmic ray mitigation approach studied here could be applied to
many instruments and detectors.

\acknowledgements
These studies are supported by the NASA Remote Exploration and
Experimentation Project (REE), which is administered at the Jet
Propulsion Laboratory under Dr. Robert Ferraro, Project Manager.

We acknowledge our colleague Craig McCreight at NASA's Ames
Research Center for his contributions in bringing all of the authors
together.

We thank the reviewer of this paper for their useful suggestions
during the referee process.

\appendix

\section{Signal-to-Noise}\label{App:SNR}

We selected Up-the-Ramp sampling for study because it provides
better signal-to-noise in what is probably the most
difficult-to-measure regime, the read-noise limit.  In the absence of
cosmic rays, Up-the-Ramp sampling provides modestly ($\sim6\%$) higher
signal-to-noise than does Fowler Sampling (Garnett \& Forrest, 1993).
The fact that an Up-the-Ramp sequence can be screened for cosmic
rays and other glitches improves this result.  Furthermore, on-the-fly
cosmic ray rejection allows longer integration times which also
improves the signal-to-noise in the faint limit.

Fowler sampling reduces the effect of read noise\footnote{We treat
read noise as random ``white noise''.} to
$\sigma^{\prime}_{r}~=~\sigma_{r}\sqrt{\frac{4}{N}}$ (for an
observation sequence consisting of $N$ samples, $N/2$ Fowler-pairs).
However, when a pixel is impacted by a cosmic ray during an
observation, the cosmic ray essentially injects infinite variance and
reduces the signal to noise to zero at that location.  If we start
with the Fowler sampling signal-to-noise function in the read-noise limit, from
Garnett \& Forrest (1993, Eqn.~6),
\begin{equation}\label{eq:SNF} SN_{F} = \frac{F T}{\sqrt{2}\sigma_r} \sqrt{\frac{\eta T}{2\cdot\delta t}}\left(1-\frac{\eta}{2}\right) = \frac{F T}{\sqrt{V_0}} \end{equation}
where $F$ is the flux of the target, $T$ is the observation time,
$\sigma_r$ is the read noise, $\eta$ is the Fowler duty cycle and $\delta
t$ is the time between sample intervals (determined by engineering or
scientific constraints on the system).  We note this
formula breaks down for relatively small numbers of samples
(i.e. $\delta t$ large with respect to $T$).  $F T$ is the signal, so
the remaining terms are the noise, which is the square-root of the
variance, $V_0$.  If we consider two cases, ``no-cosmic-ray'' and
``hit-by-cosmic-ray'' and combine the variances according to 
\begin{equation}\label{eq:VarC} V_{comb} = \frac{V_0 P_0 W_0^2 + V_1 P_1 W_1^2}{(W_0 P_0 + W_1 P_1)^2} \end{equation}
we can rewrite Equation~\ref{eq:SNF} as
\begin{equation}\label{eq:SNFC} SN_{FC} = \frac{F T}{\sqrt{V_{comb}}} = \frac{F T (W_0 P_0 + W_1 P_1)}{\sqrt{V_0 P_0 W_0^2 + V_1 P_1 W_1^2}}\end{equation}
As the weight is the inverse of the variance ($W_i = 1/V_i$),
Equation~\ref{eq:SNFC} can be rewritten as
\begin{equation}\label{eq:SNFC2} SN_{FC} = \frac{F T (\frac{P_0}{V_0} + \frac{P_1}{V_1})}{\sqrt{P_0/V_0 + P_1/V_1}} = F T \sqrt{\frac{P_0}{V_0} + \frac{P_1}{V_1}}\end{equation}
$V_0$ is the variance in the no-cosmic-ray case, taken from
Equation~\ref{eq:SNF} and $P_0$ is the probability of a pixel
surviving without a cosmic ray hit.  For simplicity, we define $1-P$
to be the probability of a pixel being hit by a cosmic ray per time
unit $\delta t$, so $P$ is the probability of ``survival'' and $P_0
= P^{T/\delta t}$.  As a cosmic ray hits injects infinite
uncertainty, the variance in the cosmic ray case $V_1 = \infty$.
Plugging in to Equation~\ref{eq:SNFC2}, we get the signal-to-noise for
Fowler sampling in the read-noise limited with cosmic rays,
Equation~\ref{eq:SNFF}:
\begin{equation}\label{eq:SNFF} SN_{FC} = F T \sqrt{\frac{P_0}{V_0} + 0} = \frac{F T}{\sqrt{2}\sigma_r}\sqrt{\frac{\eta T}{2\cdot\delta t}}\left(1-\frac{\eta}{2}\right)P^{T/(2\cdot\delta t)}\end{equation}

For a given integration time $T$ and minimum read time $\delta t$, the
maximum $SN_{FC}$ occurs with duty cycle $\eta = 2/3$.  If we plug
this back into Equation~\ref{eq:SNFF}, we get
\begin{equation}\label{eq:SNFO} SN_{F} = \frac{2}{3}\frac{F
T}{\sqrt{2}\sigma_r}\sqrt{\frac{T}{3 \delta t}}P^{T/(2\cdot\delta
t)}\end{equation} From here, it is possible to find the value of $T$
which gives the best signal-to-noise for a single observation; it
occurs at $T~=~-3\cdot\delta t/\ln(P)$.  If, however, we consider the
observation as a series of $M$ equal observations with a specific
total observation time, $T_{obs}$, the signal-to-noise for the series
is
\begin{equation}\label{eq:SNFS} SN_{FC}^{'} = \frac{F T \sqrt{M}}{3\sigma_r}\sqrt{\frac{2T}{3 \delta t}}P^{T/(2\cdot\delta t)} = \frac{F T \sqrt{T_{obs}}}{3\sigma_r\sqrt{T}}\sqrt{\frac{2T}{3\delta t}}P^{T/2\cdot\delta t}\end{equation}
If we hold $T_{obs}$ constant and find the optimum $T$, we find that
it is $T~=~-2\cdot\delta t/\ln(P)$.  In either case, it is important
to note that, in there is an optimal value for $T$, and extending the
observation beyond that time will ruin the data.

It is worth noting that the result assumes that all cosmic ray events
can be identified {\it a posteriori}.  This is not necessarily the
case, particularly when it is considered that, in the one-image case,
the fraction of pixels surviving without a cosmic ray impact is
$P^{-3/\ln(P)}=e^{-3}\approx0.05$; for the multi-image case, the
fraction of survivors is $P^{-2/\ln(P)}=e^{-2}\approx0.14$.  In both
cases, the number of ``good'' pixels is so low that separating them
from the impacted pixels will not be a trivial task.  For example, the
median operation would not be able to identify a good samples, as more
than half of the samples would be impacted by cosmic rays.  In practice,
the detector will often saturate before this limit is reached, but
this shorter integration time means that less-than-optimal
signal-to-noise will be obtained.

Up-the-Ramp sampling reduces the effect of read noise to
$\sigma^{\prime}_{r}~=~\sigma_{r}\sqrt{\frac{12}{N}}$, for N
uniformly-spaced samples with equal weighting (which is the optimal
weighting for the read-noise limited case).  When a pixel is impacted
by a cosmic ray, the Up-the-Ramp algorithm preserves the ``good'' data
for that pixel.  The exact quality of the preserved data depends on
the number of cosmic ray hits and their timing within the observation.
For example, a cosmic ray hit which just trims off the last sample in
the sequence has minimal impact compared to a cosmic ray hit that
occurs in the middle of the observation sequence.  The variance of a
Uniformly-sampled sequence with $N_i$ samples is proportional to
$1/N_i(N_i+1)(N_i-1)$.  If an Up-the-Ramp sequence is broken into
$i$ chunks by a cosmic ray, the variance becomes
\begin{equation}
V_i = V_U\frac{N(N+1)(N-1)}{\sum_{j=0}^{i}(N_j)(N_j+1)(N_j-1)}\end{equation}
  When there are zero cosmic ray
events, of course, $V_0 = V_U$.  If there is one cosmic ray event during
the sequence, the variance becomes \begin{equation}\label{eq:VarU}
V_1~=~V_U~\frac{N(N+1)(N-1)}{N_i(N_i+1)(N_i-1)+(N-N_i)(N-N_i+1)(N-N_i-1)}
\end{equation} If we assume (as is reasonable) that the cosmic ray
events are randomly distributed over time and find the expectation
value for all values of $0...N_i...N$, we find that the typical
$V_1~\approx~V_U*2$ (plus a small term in $N^{-1}$, which we will ignore
for simplicity).  If we perform a similar computation for two cosmic
ray events, we find that $V_2~\approx~V_U*10/3$ (again, plus lower-order
terms which we ignore).  In general, we find that it is possible to
find a valid result with a finite (although not necessarily pretty)
variance for any sequence broken up by cosmic ray events provided we
have at least 2 consecutive ``good'' samples (for all practical
purposes, we can ignore the situation where this is not the case).  To
simplify the following, we will simplify by considering only 3 cases:
The no-cosmic-ray case $V_0=V_U$, the one-cosmic-ray case $V_1=2*V_U$
and all multiple-cosmic-ray cases combined as one,
$V_{2+}=V_U/\epsilon^2$, where $\epsilon^2$ is a small but non-zero
number, roughly 0.3.

The Up-the-Ramp signal-to-noise function for the read-noise limited
case (Garnett \& Forrest, 1993, Eqn.~20) is \begin{equation} SN_U =
\frac{F T}{\sqrt{2}\sigma_r}\sqrt{\frac{N^2-1}{6N}}=\frac{F
T}{\sqrt{V_U}} \end{equation} We combine the variances in the 3
possible cases with the three-case equivalent to
Equation~\ref{eq:VarC}, and thus arrive at 
\begin{equation}\label{eq:SNUC} SN_{UC} = F T \left(\frac{P_0}{V_0} + \frac{P_1}{V_1} + \frac{P_{2+}}{V_{2+}}\right)^{1/2} = \frac{F T}{\sqrt{V_U}}\left(P_0 + \frac{P_1}{2} + \epsilon P_{2+}\right)^{1/2}\end{equation}
where $P_i$ is the probability of a pixel being impacted by $i$ cosmic
rays during the integration.  We note, as did Garnett \& Forrest, that
there would be no reason to limit the number of samples to anything
less than the maximum possible number, so we can set $N~=~T/\delta t$.
Using the definition of $P$ described earlier, $P_0~=~P^{T/\delta t}$,
$P_1~=~(T/\delta t)\cdot(1-P)P^{T/\delta t - 1}$ and
$P_{2+}~=~1-(P_0+P_1)$.  Putting these values back into
Equation~\ref{eq:SNUC}, we get:
\begin{equation}\label{eq:SNUCa} 
SN_{UC} = \frac{F T}{\sqrt{2}\sigma_r}\sqrt{\frac{T^2-\delta
t^2}{6T\delta t}}\left[(1-\epsilon)P^{T/\delta t} +
(1-\epsilon)\frac{T}{\delta t}(1-P)P^{T/\delta t - 1} +
\epsilon\right]^{1/2}\end{equation} If we seek the maximum value of
$SN_{UC}$ with respect to $T$, we find that $\partial(SN_{UC})/\partial T~>~0$, provided $T~\ge~\delta t$ (otherwise we'd have an integration shorter
than 1 sample time, which would be useless), $P\neq0$ and
$0<\epsilon<1$ (both of which are true by construction).  This
result applies equally whether we are considering one independent
integration or a series of observations to be combined further
downstream.  As the derivative is strictly positive, the
signal-to-noise continues to increase with as sample time increases,
although as $T~\rightarrow~\infty$, the gain in signal-to-noise
asymptotically approaches zero.  So, extending the observing time
while using Up-the-Ramp sampling with cosmic ray rejection does not
damage the data (although we might be spending time with little or no
gain).  As noted earlier for the Fowler-sampling case, there is an
optimal observing time, beyond which further observation reduces the
overall signal-to-noise.

\newpage

\end{document}